# Illustrating the relativity of simultaneity

Bernhard Rothenstein[1)], Stefan Popescu[2)] and George J. Spix[3)]

1) Politehnica University of Timisoara, Physics Department, Timisoara, Romania, bernhard_rothenstein@yahoo.com
2) Siemens AG, Erlangen, Germany, stefan.popescu@siemens.com
3) BSEE Illinois Institute of Technology, USA, gjspix@msn.com

***Abstract.*** *We present a relativistic space-time diagram that displays in true magnitudes the readings (daytimes) of two inertial reference frames clocks. One reference frame is the rest frame for one clock. This diagram shows that two events simultaneous in one reference frames are not compulsory simultaneous in the other frame. This approach has a bi-dimensional character.*

## 1. Introduction

While teaching special relativity the physics professor is in the same position as a parent who helps his little child solving a math problem. Knowing algebra the parent could solve the problem in two lines, but he may have serious problems explaining it in simple but insightful terms that are comprehensible for the child. Without knowing or applying the algebra the math professor should invent intuitive solutions for each problem despite the fact that algebra would otherwise work comfortably in these cases.

In physics the Lorentz-Einstein transformations (LET) are the algebra of special relativity. Using them correctly we can solve all the problems encountered in relativistic kinematics and avoid paradoxes at same time. It is considered that[1]: "The LET is the standard way to derive formulas for relativistic phenomena. Although the derivation of these formulas is straightforward, it is rather formal and not very transparent from the point of view of physics."

Those who use the LET should be aware of the physics behind them. Presenting them as:

$$x = \gamma(x' + \beta t) \qquad (1)$$
$$y = y' \qquad (2)$$
$$t = \gamma(t' + \frac{\beta}{c}x') \qquad (3)$$

using the well established notations $\beta = Vc^{-1}$ and $\gamma = (1-\beta^2)^{-1/2}$, physicists know that they relate the space-time coordinates of events $E(x,y,t)$ and $E'(x',y',t')$ detected from the inertial reference frames K(XOY) and respectively K'(X'O'Y'). These equations hold exactly if additionally the following conditions are fulfilled:

- The corresponding axes of the two frames are parallel to each other and the OX(O'X') axes are common.
- K' moves with constant velocity $V = \beta c$ relative to K in the positive direction of the common axes.
- Events $E$ and $E'$ take place at the same point in space, whose position is defined in K by the space coordinates (x,y) respectively in K' by the space coordinates (x',y').
- The clocks of the two reference frames are synchronized following a procedure proposed by Einstein. The symbol c in the LET represents the speed of the light signal propagating in vacuum and performing the synchronization of the distant clocks. The time coordinates $t$ and $t'$ (date times) represent the readings of clocks $C(x,y)$ and $C'(x',y')$ shortly located in front of each other at the point where the events take place.
- When the clocks of the two reference frames read a zero time ($t=t'=0$) the origins O and O' of the two frames are located at the same point in space.
- Homogeneity and isotropy of space makes that each point in space is a convenient host for the origins O and O' and that each orientation of the axes of the two reference frames is as good as all the others.
- Homogeneity of time makes that there is no preferred origin for time.

Omission of anyone of the conditions imposed above could lead to paradoxes and to fierce debates in order to eliminate or avoid them. This is the reason why we consider that the investigation of the difficulties encountered by students while learning the special relativity should start with finding out if these students are aware of the conditions imposed above[2,3].

### 2 Generating simultaneous events

When we speak about a physical quantity it is advisable to mention the observer(s) or the device(s) that detect or measure it, when and where the detection takes place and the measuring devices involved. In our paper we concentrate on the concept of simultaneity and on its relative character. A thought experiment introduced by Einstein[4] and an alternative of it[5], subsequently borrowed by countless text books and journal papers, gave rise to discussions concerning the foundational and pedagogical aspects of these experiments[6,7,8,9,10,11,12]. Analysing the different approaches to the problem of teaching the relative character of simultaneity we conclude that they either involve the somehow non-intuitive fundamental relativistic effects (time dilation and length contraction) or they perform the LET of the events involved in the experiment (from the reference frame where the experiment takes place toward the reference frame relative to which the experimental device moves). In our approach we start with some very clear definitions for the simultaneity of two distant events in an inertial reference frame:

- Two distant events are simultaneous in a given inertial reference frame if the clocks located at the points where the events take place display the same time.
- Two distant events are simultaneous in a given inertial reference frame if the light signals originating from the points where the events take place arrive simultaneously at the middle point in-between these origin points.
- Two distant clocks display the same running time after synchronization. All Einstein synchronized clocks within an inertial reference frame display the same running time.
- If a point-like light source located at the origin of an inertial reference frame starts emitting light at $t=0$ in vacuum then all the events it generates after the propagation time $= t$ further occurring on a sphere of radius $r = ct$ are simultaneous.
- Two light signals emitted simultaneously by point-like sources located at equal distances from the origin are detected simultaneously at the origin.
- Two light signals emitted at different times by two point-like sources located at different points (events $E_1(x_1, y_1, t_1)$ and $E_2(x_2, y_2, t_2)$) are received simultaneously at the origin O if:

$$t_1 = -\frac{\sqrt{x_1^2 + y_1^2}}{c} \qquad (4)$$

$$t_2 = -\frac{\sqrt{x_2^2 + y_2^2}}{c}. \qquad (5)$$

The LET take a particular form if we use them for transforming the space-time coordinates of events generated by light signals when expressed in polar coordinates $(r, \theta)$ in K and respectively $(r', \theta')$ in K'. We associate the event $E'_0(0,0,0)$ with the emission of a light signal from the origin O' at a time $t'=0$ having the same space-time coordinates in all inertial reference frames. The light signal propagates along a direction $\theta'$ and generates the event $E'(x' = ct'\cos\theta', y' = ct'\sin\theta', t')$ after a propagation time $t'$ with $ct' = r'$. When detected from K the same event is characterized by the space-time coordinates $E(x = ct\cos\theta, ct\sin\theta, t)$ with $ct = r$. If the two events represent the same event then the corresponding space-time coordinates are related by:

$$x = \gamma r'(\cos\theta' + \beta) \tag{6}$$
$$y = r'\sin\theta' \tag{7}$$
$$r = \sqrt{x^2 + y^2} = \gamma r'(1 + \beta\cos\theta') \tag{8}$$
$$t = \frac{r}{c} = \gamma t'(1 + \beta\cos\theta) \tag{9}$$
$$\cos\theta = \frac{\cos\theta' + \beta}{1 + \beta\cos\theta'} \tag{10}$$

We obtain the inverse transformations by exchanging the corresponding physical quantities in K and K' and by inverting the sign of *V*. In particular after these operations (10) reads:
$$\cos\theta' = \frac{\cos\theta - \beta}{1 - \beta\cos\theta}. \tag{11}$$
Expressing the right side of (8) and (9) as a function of $\theta$ via (11) they become:
$$r = r'\frac{\gamma^{-1}}{1 - \beta\cos\theta} \tag{12}$$
and respectively:
$$t = t'\frac{\gamma^{-1}}{1 - \beta\cos\theta} \tag{13}$$
The important conclusion is that we transform the position vector lengths $(r, r')$ and the time coordinates via the same transformation factor.

If the circle $r' = R_0$ (14) represents the geometric locus of simultaneous events in K', associated with the reception at $t' = \frac{R_0}{c}$ of the light signals emitted at $t' = 0$ from the origin O' then (12) and (13) tell us that when detected from K these events are no longer simultaneous. Each event $E'(R_0, \theta', \frac{R_0}{c})$ in K' located on the circle $r' = R_0$ has its correspondent event $E(r, \theta, t)$ in K located on the ellipse (12).

**3. Illustrating the relativity of simultaneity and the thought experiments associated with it**

Equations (12), (13) and (14) leads to a relativistic space-time diagram that enables us to find out the location of an event in K' and the location of its Einstein-Lorentz transformed correspondent in K. Figure 1 shows the diagram that displays the circle (14) and the ellipses (12) and (13) for $\beta = 0.6$.

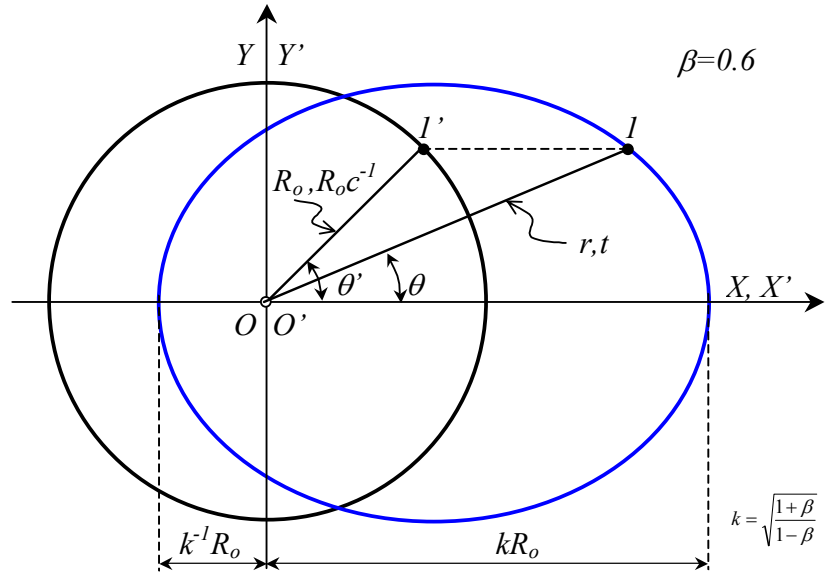

**Figure 1.** *The relativistic space-time diagram that enables us to establish the locations of the same events 1' and 1 detected from the reference frames K and K'.*

The invariance of distances measured perpendicular to the direction of relative motion tell us that the event 1 in K is the correspondent of event 1' in K'. One focus of the ellipse coincides with the centre of the circle. Let $1'(R_0, \theta'_1, t')$ and $2'(R_0, \theta'_2, t')$ be two simultaneous events in K'. The corresponding events as detected from K are $1(r_1, \theta_1, t_1)$ and $2(r_2, \theta_2, t_2)$ separated by a time interval $\Delta t = t_2 - t_1$ that depends on the location of events 1' and 2' in K' as shown in Figure 2.

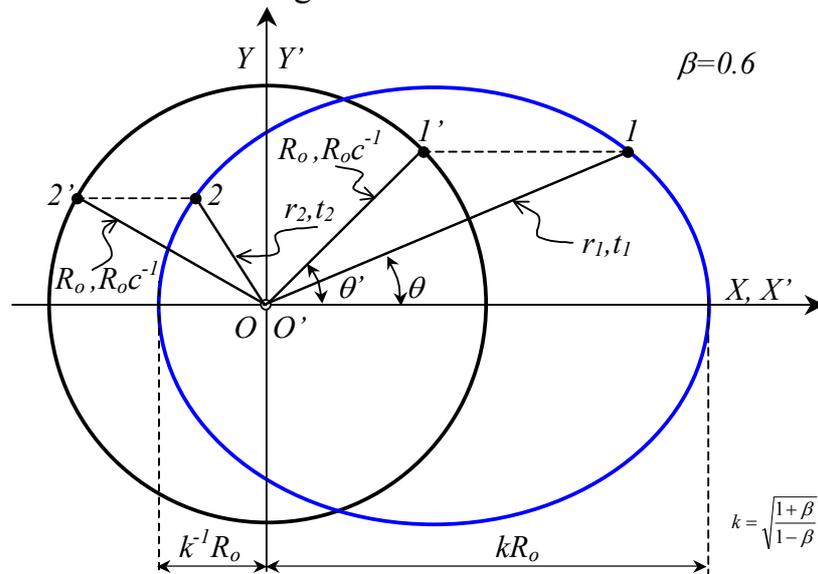

**Figure 2.** *The events 1' and 2' are simultaneous in K'. The corresponding events 1 and 2 as detected from K are no longer simultaneous.*

Consider one of the thought experiments used in the teaching the special relativity[5] sketched in Figure 3 at rest in K'.

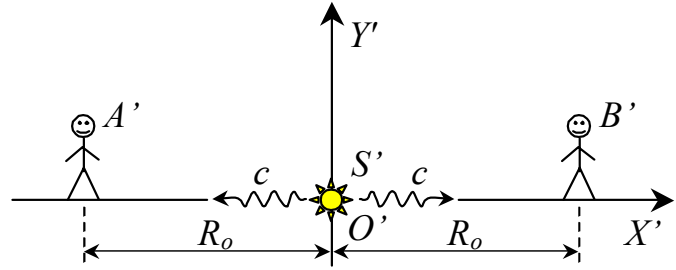

**Figure 3.** *A thought experiment used in order to illustrate the relative character of simultaneity.*

A light source is placed midway between receivers *A, A'* respectively. The distance between the two observers is $R_0$. The light source emits a flash (events O,O') towards the observers *A* and *A'*. Arriving there they generate the events $A'((-R_0,0,\frac{R_0}{c})$ and $B'(R_0,0,\frac{R_0}{c})$ which are simultaneous in K'. When detected from K the event $A(-x_A,0,t_A)$ is the correspondent of *A'* whereas the event $B(x_B,0,t_B)$ is the correspondent of event *B'*. With ($\theta = 0$) we have:

$$t_B = \frac{R_0}{c}\sqrt{\frac{1+\beta}{1-\beta}} \qquad (14)$$

and with ($\theta = \pi$) we have respectively:

$$t_A = \frac{R_0}{c}\sqrt{\frac{1-\beta}{1+\beta}}. \qquad (15)$$

As intuition tells us and the theory confirms, the event $E_A$ takes place before the event $E_B$ ($t_B > t_A$) and the two events are separated by a time interval:

$$\Delta t = t_B - t_A = \frac{2R_0}{c}\beta\gamma \qquad (16)$$

which is proportional with the distance separating the two events in the rest frame of the experimental device and with the relative velocity $V = \beta c$.

Figure 4 shows a thought experiment proposed by Einstein[4]. It is performed in two steps. We made some minor changes in the scenario, which do not alter the physics behind the experiment.

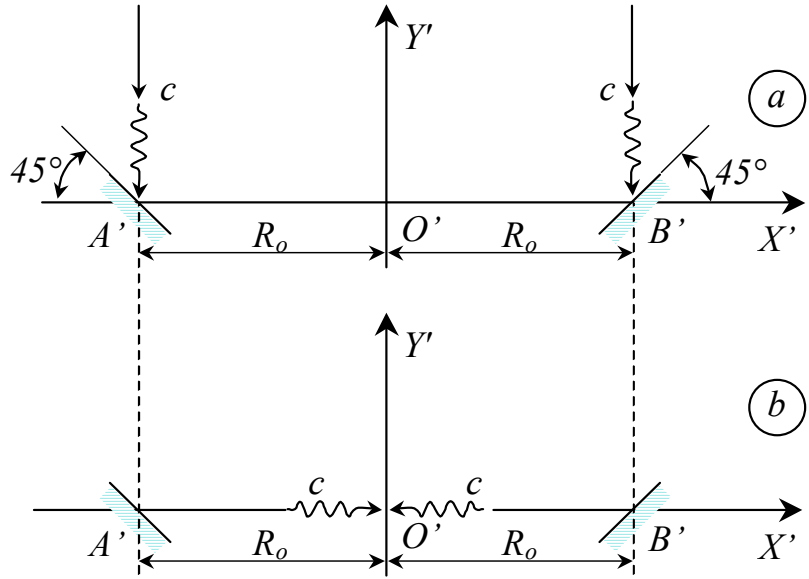

*Figure 4. Einstein's famous experiment largely used in order to illustrate the relative character of simultaneity.*

Figure 4a shows the first part of the experiment performed in the rest frame of the experimental device K'. Two mirrors A' and B' tilt at $45^0$ relative to the O'X' axis are located at equal distances $R_0$ from O'. Two rays of a plane electromagnetic wave, propagating in the negative direction of the O'Y' axis, arrive simultaneously at the location of the two mirrors generating the events $A'(-R_0, 0, -\frac{R_0}{c})$ and $B'(R_0, 0, -\frac{R_0}{c})$ respectively. Reflected by the mirrors the two rays return simultaneously to the origin O' generating the event $O'(0,0,0)$ that has the same space-time coordinates in all inertial reference frames (Figure 4b).

Detected from K the corresponding events are $O(0,0,0)$, $A(x_A, 0, t_A)$ and $B(x_B, 0, t_B)$. Figure 5 shows the relativistic space-time diagram that works in the new situation.

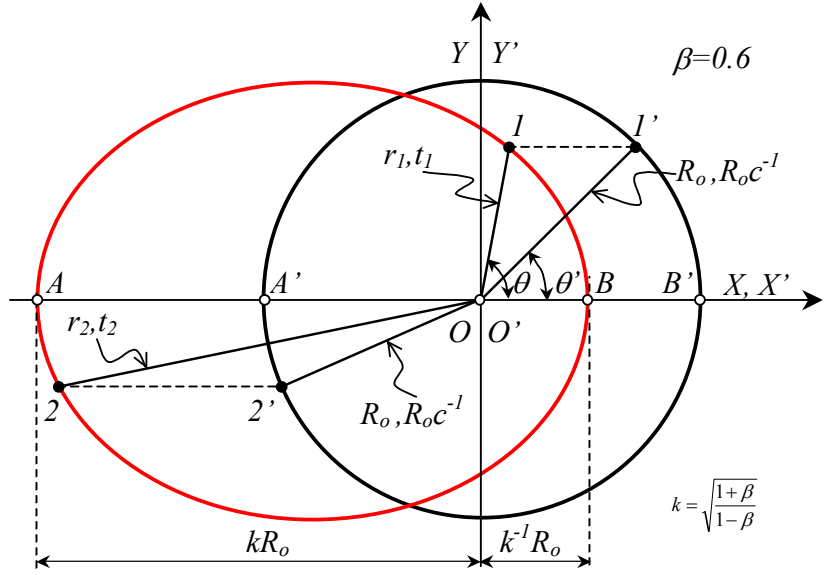

**Figure 5.** *The relativistic space-time diagram in the case of converging light signals.*

The diagram displays the geometric locus of the simultaneous events in K' (the circle $r' = -\dfrac{R_0}{c}$) and the geometric locus of the corresponding events detected from K (the ellipse $t = -\dfrac{R_0}{c}\dfrac{\gamma^{-1}}{1+\beta\cos\theta}$  (17) ).

The invariance of distances measured perpendicular to the direction of relative motion makes that event 1 corresponds to event 1', 2 corresponds to 2', A corresponds to A' and B corresponds to B'. We have:

$$t_A = -\sqrt{\dfrac{1+\beta}{1-\beta}}\dfrac{R_0}{c} \qquad (18)$$

$$t_B = -\sqrt{\dfrac{1-\beta}{1+\beta}}\dfrac{R_0}{c}. \qquad (19)$$

The time interval separating the events O and A is:

$$\Delta t_{AO} = \sqrt{\dfrac{1+\beta}{1-\beta}}\dfrac{R_0}{c} \qquad (20)$$

The time intervals separating the events O and *B* and the events *B* and *A* are respectively:

$$\Delta t_{OB} = -\sqrt{\dfrac{1-\beta}{1+\beta}}\dfrac{R_0}{c} \qquad (21)$$

$$\Delta t_{BA} = -2\gamma\beta\dfrac{R_0}{c} \qquad (22)$$

The diagram illustrates the way in which the Lorentz-Einstein transformations change in different ways the time coordinates of simultaneous events. The event associated with the return of the two rays to point O' has the same time coordinates in all inertial reference frames.

### 4. Conclusions

"Nature has unfortunately denied me the gift of being able to communicate, so that what I write is correct, but also thoroughly indigestible" says Albert Einstein about himself[13]. However his transformation equations succeeded to become a universal tool that reduces communication to zero and avoids paradoxes at same time.

The relativistic diagram that we propose here illustrates quantitatively the way in which the transformation equations differently change the time coordinates of simultaneous events in a given inertial reference frame.